# Spin transport parameters in metallic multilayers determined by ferromagnetic resonance measurements of spin-pumping*

C. T. Boone, Hans T. Nembach, Justin M. Shaw, T. J. Silva

National Institute of Standards and Technology, Boulder, CO 80305

Abstract: We measured spin-transport in nonferromagnetic (NM) metallic multilayers from the contribution to damping due to spin pumping from a ferromagnetic $Co_{90}Fe_{10}$ thin film. The multilayer stack consisted of $NM_1/NM_2/Co_{90}Fe_{10}$(2 nm)$/NM_2/NM_3$ with varying NM materials and thicknesses. Using conventional theory for one-dimensional diffusive spin transport in metals, we show that the effective damping due to spin pumping can be strongly affected by the spin transport properties of each NM in the multilayer, which permits the use of damping measurements to accurately determine the spin transport properties of the various NM layers in the full five-layer stack. We find that due to its high electrical resistivity, amorphous Ta is a poor spin conductor, in spite of a short spin-diffusion length of 1.0 nm, and that Pt is an excellent spin conductor by virtue of its low electrical resistivity and a spin diffusion length of only 0.5 nm. Spin Hall effect measurements may have underestimated the spin Hall angle in Pt by assuming a much longer spin diffusion length.

The accurate measurement and control of damping in nanomagnets, are vital for future device applications, such as magnetic random access memory (MRAM) [1], spin-torque MRAM [2], spin-torque nano-oscillators [3], racetrack memory [4], and energy-assisted magnetic recording in hard-disk drives [5,6]. Most of these applications require structures with nanoscale ferromagnets in ohmic contact with nonferromagnetic (NM) metals. A significant source of damping in such metallic multilayer (ML) structures is spin-pumping [7,8,9,10,11], whereby time-varying magnetization generates a pure spin-current into the contacting NM. Indeed, previous work has shown that spin-pumping is a major source of damping in high anisotropy CoFe/Pd multilayers [11]. While the phenomenon of spin-pumping is well established [12,9,13], the understanding of the details associated with interfaces and varying material properties throughout ML stacks still warrants further investigation to clarify remaining inconsistencies in reported data, e.g., the very different values for the enhanced spin-pumping damping due to the presence of Ta in multilayers stacks which varies from a negligible to a substantial effect [12,10]. In addition, a common approach to data analysis relies on simple models that employ an effective interfacial spin-mixing conductance to characterize MLs [9,14]. Such simplification comes at the expense of a detailed understanding of the importance that diffusive spin transport between multiple conducting NM layers can play [15]. For example, we know that non-negligible spin-accumulation due to spin-pumping in adjacent NM layers can result in a backflow of spin-current into the precessing ferromagnet (FM) [7,8]. While the details of this process can depend in a complicated way on the sample geometry, as well as the bulk and interfacial properties of any adjacent NM layers in a ML stack, the net effect is often bundled into a single phenomenological parameter: the effective spin-mixing conductance $g_{\text{eff}}^{\uparrow\downarrow}$ [7,9,11,16].

Spin-pumping from a uniformly precessing ferromagnet (FM) causes an increase in the effective Gilbert damping $\alpha$ [7,8,17]. The spin-current pumped from the time-varying magnetization into an adjacent NM layer propagates diffusively, subject to the operative boundary conditions in the NM system, including any NM interfaces in the case of a ML stack. The net spin-current exiting the FM is proportional to the gradient of the spin accumulation at the FM/NM interface [7]. Given that the resultant spin accumulation is a sensitive function of the electrical conductivities, spin-diffusion lengths, and interfacial spin-flip-scattering of the various NM layers in the ML stack, accurate determination of both the bulk and interfacial spin-transport properties of the constituent materials is critical for the optimized design of spintronic devices that utilize ML stacks of NM layers.

In this work, we experimentally assess the effects of multiple NM layers on the spin-pumping contribution to damping of the single FM layer in the stack, and model the enhanced damping with conventional diffusive spin-current theory for one-dimensional transport of non-collinear spins. We find that the presence of a material with a short spin-diffusion length in a multilayer stack does not necessarily guarantee a large effective spin-mixing conductance or large enhanced damping. In particular, a large mismatch in spin-conductivity between two layers in a multilayer stack can generate a substantial spin accumulation in a relatively weak "spin scatterer" (i.e., long spin diffusion length), even though it is in direct contact with a strong "spin scatterer" (i.e., short spin diffusion length).

The standard theory for diffusive spin-transport is used to quantitatively analyze the additional damping due to spin-pumping [7,8]. Spins that are pumped from the FM diffuse into the NM multilayer, leading to a steady-state, non-collinear, spin accumulation throughout the structure. For uniform, circular, out-of-plane precession, the pumped spins flow in a direction perpendicular to the interface, with a polarization perpendicular to the instantaneous magnetization vector and parallel to the FM/NM interface to first order in the precession amplitude. Given that the ferromagnetic resonance (FMR) frequencies for these measurements (5-30 GHz) are much slower than the electron relaxation rates in the NM (>200 GHz), the system reduces to a time-independent, one-dimensional problem defined by the spin-potential $\vec{\mu}_s(x)$ and spin-current $\vec{j}_x^s(x)$, where the spin-polarization vectors are parallel to $-\hat{m}\times(d\hat{m}/dt)$, and the subscript for the spin-current refers to the flow along the x-axis, perpendicular to the layers. The FM/NM interface lies at $x=0$, and the NM stack extends in the half plane $x>0$. (Given the symmetry of our sample system, we will only solve the spin-diffusion problem for one of the FM/NM interfaces, and then multiply our results for enhanced damping by a factor of two.) The steady-state transverse spin-potential is given by the solution to the spin-diffusion equation [8,18],

$$\frac{d^2\vec{\mu}^s}{dx^2} = \frac{1}{\lambda^2}\vec{\mu}^s, \qquad (1)$$

where $\lambda$ is the spin-diffusion length. Eq.(1) is subject to the boundary condition that both the spin-potential and the spin-current, defined by

$$\vec{j}_x^s = -\frac{\hbar\sigma}{2e^2}\frac{d\vec{\mu}^s}{dx}, \qquad (2)$$

($\sigma$ is the electrical conductivity), are continuous across all NM interfaces, where $\vec{j}_x^s = 0$ at the outermost NM/air interface, and the spin-current and spin-potential satisfy the spin-pumping/spin-torque source term at the FM/NM interface [7,8], given by

$$\vec{j}_x^s(x=0) = \frac{\hbar\,\mathrm{Re}[G_{\uparrow\downarrow}]}{2e^2}\left[\hbar\frac{d\hat{m}}{dt}\times\hat{m} - 2\vec{\mu}^s(x=0)\right] \qquad (3)$$

where $e$ is the electron charge, $\hbar$ is the reduced Planck's constant, and $G_{\uparrow\downarrow}$ is the spin-mixing conductivity with units of $\Omega^{-1}$ m$^{-2}$. The conversion between the $G_{\uparrow\downarrow}$ used here and the oft-used alternative form, referred to as the the spin-mixing conductance $g_{\uparrow\downarrow}$, with units of m$^{-2}$, is the von Klitzing resistance of $\cong 26$ k$\Omega$, i.e., $g_{\uparrow\downarrow} \doteq (h/e^2) G_{\uparrow\downarrow}$. We refer to G as the spin mixing conductivity, as opposed to conductance, because it relates a spin-current density to a potential drop. In the formulation of Eq. (3), the spin accumulation in the NM is assumed to be purely transverse, i.e., $\vec{\mu}^s \cdot \hat{m} \cong 0$, and the transverse spin-potential in the FM is zero such that $\vec{\mu}^s(x=0)$ is equal to the interfacial spin-potential drop at the FM/NM interface.) Rewriting Eq. (3) in terms of the torque acting on the FM due to spin current at the FM/NM interface, the spin-pumping component of the damping $\Delta\alpha$ for each FM/NM interface is [7,8]

$$\Delta\alpha = \frac{|\gamma|\hbar^2}{2M_s \delta e^2} \frac{\text{Re}[G_{\uparrow\downarrow}]}{\left[1+\frac{\text{Re}[G_{\uparrow\downarrow}]}{G_{ext}}\right]}, \qquad (4)$$

where $G_{ext} \doteq [2e^2/\hbar] [|\vec{j}_x^s(x=0)|/|2\vec{\mu}^s(x=0)|]$ is the effective one-dimensional spin-conductance of the NM stack, which acts as a parallel spin-conductor (or series spin-resistor) to the spin-mixing conductance, $|\gamma|$ is the gyromagnetic ratio assuming the experimentally-determined Landé g-factor g=2.15, $M_s$ is the saturation magnetization, with $\mu_0 M_s = 1.8$ T for Co-Fe [11], and $\delta$ is the ferromagnetic layer thickness. The ratio $\text{Re}[G_{\uparrow\downarrow}]/G_{ext}$ in Eq. (4) is usually referred to as the spin-current "backflow factor" [7], and the factor $\text{Re}[G_{\uparrow\downarrow}]/[1+[\text{Re}[G_{\uparrow\downarrow}]/G_{ext}]] < \text{Re}[G_{\uparrow\downarrow}]$ is sometimes referred to as the "effective" spin-mixing conductance. We note that each layer in the NM stack cannot be treated as a separate parallel spin-resistor as part of a simplified lumped-circuit analysis because the steady-state spin-transport properties of the aggregate ML stack depend non-trivially on the thickness of each NM layer, as will be shown below. For purely diffusive systems, we generally assume [7,8] that $\text{Im}[G_{\uparrow\downarrow}] \ll \text{Re}[G_{\uparrow\downarrow}]$. Thus, $\text{Re}[G_{\uparrow\downarrow}]$ will be replaced with $G_{\uparrow\downarrow}$ for the rest of our analysis.

The homogeneous solution to Eq. (1) in NM layer $i$ is

$$\mu_i^s(x) = A_i \exp\left(\frac{x}{\lambda_i}\right) + B_i \exp\left(-\frac{x}{\lambda_i}\right), \tag{5}$$

where $\lambda_i$ is the spin diffusion length in layer $i$, and $(A_i, B_i)$ are coefficients to be solved by use of the boundary conditions. For a FM/NM$_1$/NM$_2$ multilayer,

$$G_{ext} = \frac{G_1^{NM}}{2} \frac{\left[G_1^{NM} \coth\left(\frac{L_2}{\lambda_2}\right) + G_2^{NM} \coth\left(\frac{L_1}{\lambda_1}\right)\right]}{\left[G_1^{NM} \coth\left(\frac{L_1}{\lambda_1}\right) \coth\left(\frac{L_2}{\lambda_2}\right) + G_2^{NM}\right]}, \tag{6}$$

where $G_i^{NM} \doteq \sigma_i/\lambda_i$, $\sigma_i$ and $L_i$ are the electrical conductivity and the thickness of the $i$th layer, respectively. $G_i^{NM}$ can be regarded as an effective spin-conductivity for the $i$th layer. In the limit where the second layer is much thicker than its spin diffusion length ($L_2 \gg \lambda_2$), Eq. (6) reduces to

$$G_{ext} = \frac{G_1^{NM}}{2} \frac{\left[G_1^{NM} \tanh\left(\frac{L_1}{\lambda_1}\right) + G_2^{NM}\right]}{\left[G_1^{NM} + G_2^{NM} \tanh\left(\frac{L_1}{\lambda_1}\right)\right]}. \qquad (7)$$

Illustrating the spin accumulation expected for the kinds of metallic multilayers considered here, Fig. 1(a) shows a plot of $|\vec{\mu}^s|$ as a function of position for the three cases of $\sigma_2 = 3\sigma_1$, $\sigma_2 = \sigma_1$, and $\sigma_2 = 0.06\sigma_1$, with $2L_1 = L_2 = \lambda_1 = \lambda_2$, calculated by use of Eq. (6). In the limit of $G_{ext} \ll G_{\uparrow\downarrow}$, we see from Eqs. (2) and (4) that $\Delta\alpha \propto G_{ext} \propto \left[d\mu^s/dx\right]_{x=0}/\mu^s(x=0)$. Thus $\Delta\alpha$ can be strongly affected by changes in the electrical conductivity of the outermost NM layer, even though the spin diffusion lengths of both NM layers are identical.

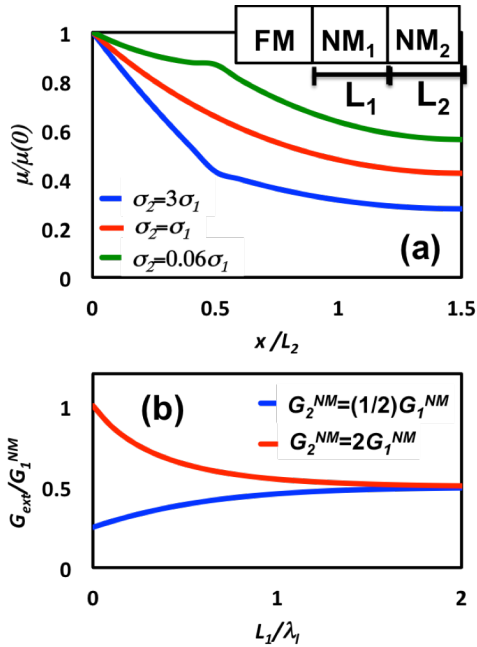

Figure 1: (a) Theoretically predicted spin accumulation as a function of distance from the interface for a bilayer with thicknesses L$_1$=0.5L$_2$, for a differing conductivities in layer two: σ$_2$=3σ$_1$ (blue), σ$_2$=σ$_1$ (red) and σ$_2$=0.06σ$_1$. Simply changing the electrical conductivity of Layer 2 can greatly affect the reabsorbed spin current due to accumulation at the ferromagnet boundary. Inset shows a schematic of the trilayer structure. (b) $G_{ext}/G_1^{NM}$ as a function of $L_1$ for $G_2^{NM} = (1/2)G_1^{NM}$ (blue) and $G_2^{NM} = 2G_1^{NM}$ (red), as calculated by use of Eq. (7). As expected by inspection of Eq. (7), $G_{ext} = G_2^{NM}/2$ for $L_1 = 0$, and asymptotically approaches $G_1^{NM}/2$ with increasing $L_1$.

Fig. 1(b) shows a plot of $G_{ext}$ vs. $L_1/\lambda_1$ calculated from Eq. (7) for different relative spin conductivities in the two layers. This example demonstrates how $G_{ext}$ can either be an increasing or decreasing function of layer thickness, with a strong dependence on the spin-conductivity of the second NM layer. Thus, by measuring $\Delta\alpha$ as a function of $L_1$, both $\lambda_1$ and the dimensionless ratio $\left[\sigma_1\lambda_2\right]/\left[\sigma_2\lambda_1\right]$ can be determined via nonlinear fitting. Additional measurements of the conductivities for the two NM layers then allow for determination of $\lambda_2$.

We investigated spin-pumping into metallic MLs by depositing stacks of Ta(3 nm)/ NM($L_{NM}$)/Co$_{90}$Fe$_{10}$(2 nm)/NM($L_{NM}$)/ Ta(3 nm), as shown schematically in Fig. 2(a), where the NM is either Pd or Cu. (Subsequently, we will refer to the alloy Co$_{90}$Fe$_{10}$ with the abbreviation "Co-Fe".) The samples were magnetron sputter-deposited in a chamber with a base pressure of $1.3\times10^{-7}$ Pascal ($10^{-9}$ Torr). The thicknesses were calibrated by means of X-ray reflectometry with an uncertainty of approximately ±5%. The high symmetry of this structure allows for simplified characterization of spin-transport because the contributions of both Co-Fe interfaces to the spin-pumping component of damping are presumably identical. Figure 2(b) shows a schematic of spin pumping from the driven, precessing magnetization of the Co-Fe layer, where spins of opposite orientation are emitted and absorbed. The effective damping is measured by use of an FMR spectrometer that employs a vector network analyzer (VNA) [11,19]. The damping parameter $\alpha$

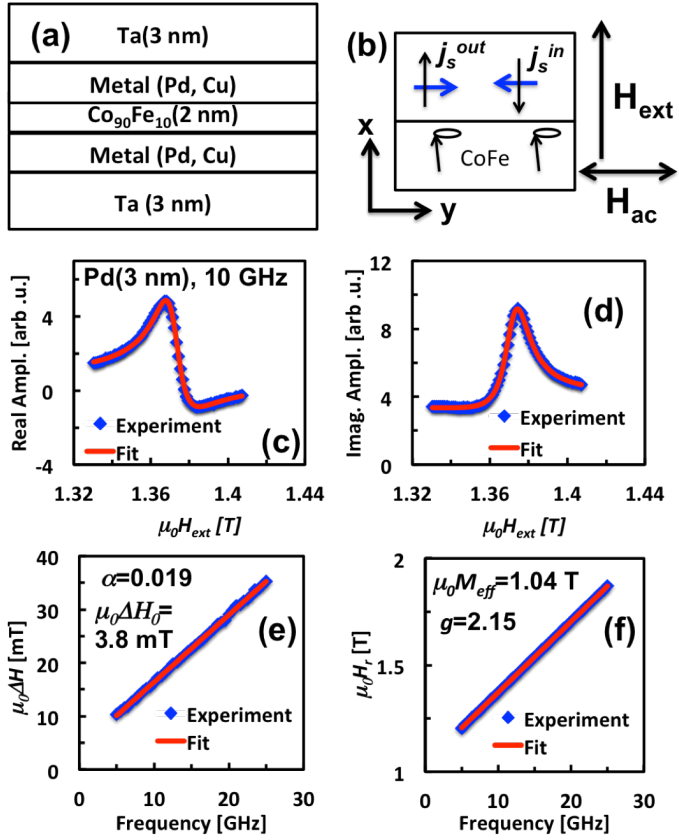

Figure 2: (a) Schematic of the multilayer stacks for studying spin-pumping. (b) Schematic of the spin-pumping phenomenon. An oscillating field $H_{ac}$ excites precession of the CoFe magnetization, which induces a steady-state spin-accumulation and spin-current in the NM layers. (c) and (d) Representative plots of the real and imaginary parts of the FMR spectrum (blue points) with fits to the Polder susceptibility (red lines). (e) $\Delta H$ versus Frequency (blue points) with linear fit (red line), yielding $\Delta H_0$ and $\alpha$. (f) Resonance field versus frequency (blue points) with fit to the Kittel equation (red line), yielding $M_{eff}$ and $g$.

is extracted from the slope of the field linewidth $\Delta H$ versus frequency $f$ [20,21]. Representative spectra and simultaneous fits to the real and imaginary part of the susceptibility for the ML with NM = Pd, $L_{Pd}$ = 3 nm, measured at 10 GHz, are shown in Figs. 2(c)-(d), while Figs. 2(e)–(f) are plots of $\Delta H$ and the resonance field $H_r$ versus $f$, along with linear fits to the data. We clearly see the linear dependence of

$\Delta H$ on $f$, which also permits explicit separation of $\alpha$ from any inhomogeneous contribution to the linewidth.

Fig. 3(a) shows the measured $\alpha$ as a function of Pd thickness $L_{Pd}$. For $L_{Pd} > 1$ nm, $\alpha$ increases by a factor of two with increasing $L_{Pd}$, with a characteristic length scale of 2-3 nm, indicating that the spin-impedance of Ta is much smaller than that of Pd (Fig. 1(b)). We measured the conductivity of our Ta to be $\sigma_{Ta} = 4.1 \times 10^5$ $\Omega^{-1}$ m$^{-1}$, and we assumed a bulk conductivity of $\sigma_{Pd} = 9.5 \times 10^6$ $\Omega^{-1}$ m$^{-1}$ [22]. The low conductivity of the Ta is consistent with X-ray diffraction data that indicate an amorphous structure. To be consistent with previously reported data for similar structures with a fixed Ta(3 nm)/Pd(3 nm) seed layer with no Ta on top, the data in both Fig. 3(a) and that of Fig. 5 in Ref [11] were simultaneously fitted with Eqs. (4) and (6). The fitting results are $\lambda_{Pd} = 2.6 \pm 0.1$ nm, $\lambda_{Ta} = 1.0 \pm 0.2$ nm, and $G_{\uparrow\downarrow}^{CoFe} = 1.26 \pm 0.17 \times 10^{15}$ $\Omega^{-1}$ m$^{-2}$, as summarized in Table 1. In all of the fits, $\alpha = \alpha_0 + \Delta\alpha$, with an intrinsic damping value of $\alpha_0 = 0.00462 \pm 0.00004$, as established by previous measurements [11]. The imprecision in the fitted values is primarily attributable to the ±5 % uncertainty in the film thickness. Additional physical effects not contained in Eq. (6) were also accounted for. Because approximately 1 nm of Pd is ferromagnetically polarized when in direct contact with Co-Fe [11], we subtracted $t_0 = 1$ nm from $L_{Pd}$, such that $G_{ext}(L_{Pd}) \to G_{ext}(L_{Pd} + t_0)$, while also adding 0.2 nm to the total effective Co$_{90}$Fe$_{10}$ thickness $\delta$ in Eq. (4) to account for the magnetization of the polarized Pd equal to one-tenth that of Co-Fe. Nevertheless, the data clearly demonstrate how the poor conductivity of the Ta layer can lead to weak spin pumping when the intermediate Pd layer is thinner than the spin diffusion length, even though the spin diffusion length of Ta is almost one-third that of Pd.

Previously reported values of $G_{\uparrow\downarrow}^{\text{CoFe/Pd}} = (4.15 \pm 0.50) \times 10^{14}$ Ω$^{-1}$ m$^{-2}$ and $\lambda_{\text{Pd}} = 8.6 \pm 1.0$ nm were also derived from spin-pumping measurements with similar multilayer structures [11]. While these earlier results compared favorably with values reported by Foros et al. [14] and Mizukami et al. [12], the analysis used in both Refs. [11] and [14] was based upon the assumption that $\Delta\alpha$ exhibits a simple exponential dependence on $L_{\text{Pd}}/\lambda_{\text{Pd}}$, which is now known to be an oversimplification. The quantitative analysis presented here, based upon Eqs. (4) and (6), takes into account the well-known proximity-induced polarization of Pd and has greatly improved the accuracy of the fitted parameters.

| NM | $G_{\uparrow\downarrow}^{\text{CoFe/NM}}$ (10$^{15}$ Ω$^{-1}$ m$^{-2}$) | $\lambda$ (nm) | $\sigma$ (Ω$^{-1}$ m$^{-1}$) | $\sigma/\lambda$ (10$^{15}$ Ω$^{-1}$ m$^{-2}$) |
|---|---|---|---|---|
| Pd | 1.26±0.17 | 2.6±1.2 | 9.5× 10$^6$ [22] | 3.65±0.49 |
| Ta | N/A | 1.0±0.18 | 4.1× 10$^5$ | 0.41±0.074 |
| Cu | >2.6 | 170 [23] | 5.8×10$^7$ [22] | 0.32 |
| Nb | N/A | 8.0±2.5 | 6.6×10$^6$ [22] | 0.26 |
| Pt | N/A | 0.5±0.3 | 9.4×10$^6$ [22] | 19 |

Table 1. Summary of fitted and tabulated parameters. Citations are provided for tabulated parameters that were used as part of the fitting process.

We speculate that the large value for $\alpha$ for $L_{Pd} = 0$ (Ta (3 nm)/ Co-Fe (2nm)/Ta (3 nm)) is the consequence of a magnetic "dead layer" at the $Co_{90}Fe_{10}$/Ta interface that further reduces the effective thickness of the Co-Fe [24]. Enhancement of $G_{\uparrow\downarrow}$ for the Co-Fe/Ta interface relative to that of the Co-Fe/Pd interface cannot account for the large $\alpha$ value because it is already the case that $G_{Ta} \ll G_{\uparrow\downarrow}^{CoFe/Pd}$. Eq. (4) shows that the spin-pumping contribution to $\alpha$ is insensitive to $G_{\uparrow\downarrow}$ in the limit of $G_{ext} \ll G_{\uparrow\downarrow}$. Based on the measured value of $\alpha$ when the Co-Fe and Ta are in direct contact, and assuming $G_{\uparrow\downarrow}^{CoFe/Ta} \cong G_{\uparrow\downarrow}^{CoFe/Pd}$, the dead layer thickness is estimated to be ≈ 0.45 nm per interface. This is in reasonable agreement with a previously reported value of 0.66 nm for a sputtered Co/Ta interface [25].

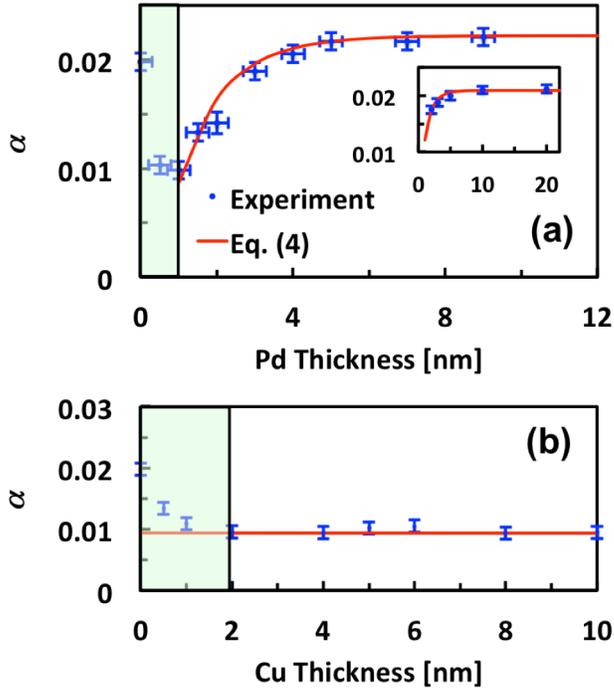

Fig. 3: (a) Measured $\alpha$ (blue points) vs. Pd thickness, with self-consistent fits (red lines) to Eq. (4), for the symmetric multilayer. Inset shows the simultaneous fit to data contained in Fig. 5 of Ref. [11]. (b) Same as (a) but with Cu instead of Pd. Shaded regions indicate regions below $t_0$, where, due to polarization of the normal metal and inconsistent growth the theory breaks down. Only data outside of the shaded regions in (a) and (b) are fitted to Eq. (4).

As a control study, we repeated our measurements with similar samples, but with NM = Cu. Cu has very similar spin conductivity to that of Ta, as indicated in Table **1**, even though Cu has an exceptionally long spin diffusion length. Fig. 3(b) shows measured values of $\alpha$ as a function of Cu thickness $L_{Cu}$. For all of the measured samples, $L_{Cu} \ll \lambda_{Cu}$. Over the entire range of 1 nm $< L_{Cu} \leq$ 10 nm, $\alpha$ is effectively constant, with a value of $\cong 0.01$, nearly identical to the minimum value obtained for the samples with NM = Pd, $L_{Pd} = 1$ nm. This is consistent with both the fact that $\sigma/\lambda$ is very similar for Cu and Ta, and that $G_{\uparrow\downarrow} \gg G_{Cu,Ta}$ for both the Co-Fe/Cu and Co-Fe/Pd interfaces. We attribute the enhanced damping for $0 < L_{Cu} < 1.5$ nm to discontinuous Cu growth at these small thicknesses. By use of the minimum value of $\Delta\alpha$ for $L_{Cu} \geq 2$ nm, and the previously measured value $\lambda_{Cu} \cong 170$ nm [26], fitting of the data yields $G_{\uparrow\downarrow}^{CoFe/Cu} > 2.6 \times 10^{15}$ $\Omega^{-1}$ m$^{-2}$. Establishment of an upper bound for $G_{\uparrow\downarrow}^{CoFe/Cu}$ was not possible, given the finite precision of the measurements and the relative insensitivity predicted by Eq. (4) when $G_{\uparrow\downarrow}^{CoFe/Cu} \gg G_{Cu}$. Nevertheless, by use of the lower bound for $G_{\uparrow\downarrow}^{CoFe/Cu}$ and the other fitted parameters in Table 1, and with the assumption $G_{\uparrow\downarrow}^{CoFe/Cu} \cong G_{\uparrow\downarrow}^{NiFe/Cu}$, we were able to quantitatively reproduce the dependence of $\Delta\alpha$ on $L_{Cu}$ as reported in Ref. [10] for multilayers with the structure $Ni_{80}Fe_{20}$ (5 nm)/Cu ($L_{Cu}$)/Ta (5 nm) and $Ni_{80}Fe_{20}$ (5 nm)/Cu ($L_{Cu}$), where $L_{Cu}$ was varied between < 5 nm to > 1 μm.

As an additional control measurement, we grew samples with Ni layers of thickness $\delta_{Ni}$ between the Pd and Ta (see diagram in Fig. 4). The FMR conditions for Ni are sufficiently different from those for Co-Fe that the Ni magnetization is approximately static over the Co-Fe resonance peak. The Ni is an absorber of transverse spin ($\vec{\mu}^s \cdot \vec{M}_{Ni} \cong 0$) via the spin-torque effect, as described quantitatively by the second term in Eq. (3). The efficiency of the spin absorption by Ni is proportional to both the spin-mixing conductance of the Ni/Pd interface $G_{\uparrow\downarrow}^{Ni/Pd}$ and a factor of the approximate form $1 - \text{sech}(-\delta_{Ni}/\lambda_{Ni})$, where $\lambda_{Ni} \cong 1$ nm is the transverse spin dephasing length in Ni [27]. In the limit of $L_{Pd} \ll \lambda_{Pd}$ and $\delta_{Ni} \gg \lambda_{Ni}$, the damping due to spin pumping should simply be that predicted by Eq. (4), but with $G_{ext} = G_{\uparrow\downarrow}^{Ni/Pd}$ due to the fact that the net transverse spin current pumped out of the Co-Fe relaxes entirely at the Ni/Pd interface. Assuming that $G_{\uparrow\downarrow}^{Ni/Pd} \cong G_{\uparrow\downarrow}^{CoFe/Pd} \gg G_{Pd}$, as expected by theory [14] and prior experiments [13], the inclusion of a sufficiently thick Ni layer should significantly enhance $\Delta\alpha$ for $L_{Pd} < \lambda_{Pd}$. This is seen in Fig. 4(a), where we present the data for $\Delta\alpha$ with varying $L_{Pd}$ and several values of $\delta_{Ni}$. Thus, the measured dependence of $\Delta\alpha$ on $\delta_{Ni}$ is consistent with the theory of spin-pumping/spin-torque summarized in Eq. (3). Solving for the spin-mixing conductance using the value of $\alpha$ for $L_{Pd} = 1$ nm and $\delta_{Ni} = 2.5$ nm, with the substitution $G_{ext} = G_{\uparrow\downarrow}^{Ni/Pd}$ in Eq. (4), we obtain $G_{\uparrow\downarrow}^{Ni/Pd} = (4.7 \pm 0.7) \times 10^{15}$ $\Omega^{-1}$ m$^{-2}$.

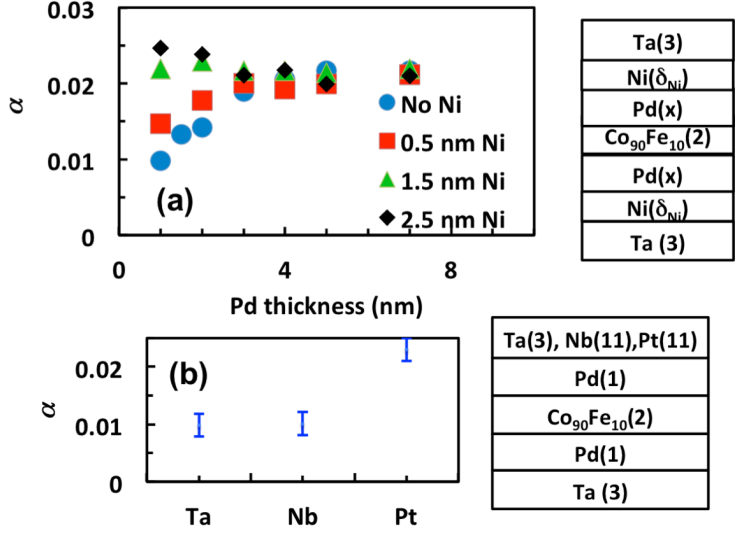

Figure 4: (a) $\alpha$ vs $L_{Pd}$ with Ni layer of thickness $\delta_{Ni}$ inserted between Pd and Ta, with schematic showing the layer structure. Ni absorbs the pumped spin-current via spin-torque, causing an increase in $\Delta\alpha$. (b) $\alpha$ for different NM capping materials of Ta, Nb, and Pt, with schematic showing the layer structure. The Pt capping layer is a good spin-conductor, whereas the Ta and Nb capping layers are poor spin-conductors.

We also investigated whether the overall trend of small/large $\Delta\alpha$ and small/large $G_{NM}$ for thin Pd was reproducible for other NM capping layers. For this final study, multilayers were grown consisting of Ta (3 nm)/Pd (1 nm)/ CoFe (2 nm)/Pd (1 nm)/NM$_{cap}$ (11nm), where NM$_{cap}$ = Nb or Pt. The data are shown in Fig. 4(b). The contribution of the bottom interface to $\Delta\alpha$ was calculated by use of the previously determined fitting parameters in Table 1. Any additional contribution to $\Delta\alpha$ by the upper Co-Fe/Pd interface was fitted to the data by use of Eq. (4). While Nb and Pt have similar conductivities to that of Pd, others have reported that they have very different values for the spin diffusion length [28,29], although there is some disagreement in the literature on the value of $\lambda_{Pt}$, which varies between 14 nm [16,30] to 1.4 nm [29]. Our data are consistent with the notion that Nb is a poor spin conductor, with $G_{Nb} \cong G_{Ta}$, whereas Pt is a good spin conductor, with $G_{Pt} \cong G_{Pd}$. As such, these data would indeed appear to reinforce the contention in Ref. [25] that $\lambda_{Pt} \cong 1$ nm. Solving for the spin diffusion length with Eq. (4), using the bulk values of the conductivity listed in Table **1**, we obtain $\lambda_{Pt} = (0.5 \pm 0.3)$ nm and $\lambda_{Nb} = (8.0 \pm 0.3)$ nm.

Note that the spin-impedance for a given element could be highly variable, depending on the phase of the material. The sputtered amorphous Ta used for these samples is a poor conductor, thereby resulting in a relatively poor spin-conductivity. Similarly, $\beta$-Ta, which was recently reported to exhibit a giant spin Hall angle [31], is also a poor conductor. However, $\alpha$-Ta is a much better conductor, which implies a much larger spin-conductivity. Thus, it is possible that the very different values for the spin-pumping effect in multilayers that utilize Ta seed and/or cap layers could possibly result from sample growth conditions that produce different phases of Ta.

In summary, we measured the spin-pumping efficiency of 2 nm Co-Fe for sandwich structures with encapsulating multilayers consisting of a variety of NM and FM layers. Interpretation of the data was facilitated by conventional diffusive spin transport theory. By self-consistent fitting of the data for samples with varying NM and FM layer thicknesses, we could accurately determine the intrinsic spin-mixing conductance of both Co-Fe/Pd and Ni/Pd interfaces, as well as the spin diffusion lengths for a variety of normal metals. We rarely found that the effective spin-mixing conductance was a good approximation for the intrinsic spin-mixing conductance. Instead, we generally needed to properly account for the details of diffusive spin-current flow between the various interfaces to accurately fit the data. The values for $G_{\uparrow\downarrow}^{\text{CoFe/Pd}}$ and $\lambda_{\text{Pd}}$ reported here should prove valuable in the successful modeling of damping behavior in more complicated systems, such as perpendicular anisotropy multilayers for spintronics applications.

[1] ITRS, "International Technology Roadmap for Semiconductors 2011, Emerging Research Devices" (2011)
[2] J. A. Katine and E. E. Fullerton, *J. Magn. Magn. Mater.* **320** (2008) 1217–1226
[3] S. I. Kiselev, et al., Phys. Rev. Lett. **93**, 036601 (2004).
[4] A. Makarov, V. Sverdlov, and S. Selberherr, *Microelectronics Reliability*, Vol. 52 Issue 4 628-634 (2012)
[5] Yiming Wang, Yuhui Tang and Jian-Gang Zhu, *J. Appl. Phys.* **105**, 07B902 (2009)
[6] C. T. Boone, J. A. Katine, E. E. Marinero, S. Pisana, and B. D. Terris, *IEEE Magn. Lett.* Vol. 3 #3500104 (2012)
[7] Yaroslav Tserkovnyak, Arne Brataas and Gerrit E. W. Bauer, *Phys. Rev. B* **66,** 224403 (2002)
[8] M. L. Polianski and P. W. Brouwer, *Phys. Rev. Lett.* **92**, 026602 (2004)
[9] O. Mosendz, J. E. Pearson, F. Y. Fradin, G. E. W. Bauer, S. D. Bader, and A. Hoffmann, *Phys. Rev. Lett.* **104**, 046601 (2010)
[10] Th. Gerrits, M. L. Schneider, and T. J. Silva, *J. App. Phys.* **99,** 023901 (2006)
[11] J. Shaw, H. Nembach, and T. J. Silva, *Phys. Rev. B* **85**, 054412 (2012)
[12] S. Mizukami, Y. Ando, and T. Miyazaki, *J. Magn. Magn. Mater.* 226-230 (2001) 1640-1642
[13] F. D. Czeschka, L. Dreher, M. S. Brandt, M. Weiler, M. Althammer, I.-M. Imort, G. Reiss, A. Thomas, W. Schoch, W. Limmer, H. Huebl, R. Gross, and S. T. B. Goennenwein, *Phys. Rev. Lett.* **107**, 046601 (2011)
[14] J. Foros, G. Woltersdorf, B. Heinrich, and A. Brataas, *J. Appl. Phys.* **97**, 10A714 (2005).
[15] M. Morota, Y. Niimi, K. Ohnishi, D. H. Wei, T. Tanaka, H. Kontani, T. Kimura, and Y. Otani, *Phys. Rev. B* **83**, 174405 (2011).
[16] O. Mosendz, J. E. Pearson, F. Y. Fradin, G. E. W. Bauer, S. D. Bader, and A. Hoffmann, *Phys. Rev. Lett.* **104**, 046601 (2010).
[17] Yaroslav Tserkovnyak, Arne Brataas and Gerrit E. W. Bauer, *Phys. Rev. Lett.* **88**, 117601 (2002)
[18] S. Takahashi and S. Maekawa, *Phys. Rev. B* **67**, 052409 (2003)
[19] H. T. Nembach, T. J. Silva, J. M. Shaw, M. L. Schneider, M. J. Carey, S. Maat, and J. R. Childress, *Phys. Rev. B* **84**, 054424 (2011)
[20] M. Farle, *Rep. Prog. Phys.* **61**, 755 (1998)
[21] R. D. McMichael, D. J. Twisselman, and Andrew Kunz, *Phys. Rev. Lett.* **90**, 227601 (2003)
[22] http://www.webelements.com

[23] J. Bass and W. P. Pratt, *J. Phys. Cond Matt.* **19**, 183201 (2007)
[24] D. Markó, T. Strache, K. Lenz, J. Fassbender, and R. Kaltofen, *Appl. Phys. Lett.* **96**, 022503 (2010)
[25] H. Lefakis, M. Benaissa, P. Humbert, V.S. Speriosu, J. Werckmann, B.A. Gurney, *J. Magn. Magn. Mater.* **154**, 1, 17-23 (1996).
[26] F. J. Albert, N. C. Emley, E. B. Myers, D. C. Ralph, and R. A. Buhrman, *Phys. Rev. Lett.* **89**, 226802 (2002).
[27] M. D. Stiles and J. Miltat, "Spin transfer torque and dynamics," in <u>Spin Dynamics in Confined Magnetic Structures III</u>, B. Hillebrands and A. Thiaville, eds. (Springer-Verlag, 2006).
[28] Wanjun Park, David V. Baxter, S. Steenwyk, I. Moraru, W. P. Pratt, Jr., and J. Bass, *Phys. Rev. B* **62**, 1178 (2000).
[29] L. Liu, R. A. Buhrman, and D. C. Ralph, arXiv:1111.3702 (2012).
[30] H. Kurt, R. Loloee, K. Eid, W. P. Pratt, and J. Bass, *Appl. Phys. Lett.* **81**, 4787 (2002)
[31] L. Liu, Chi-Feng Pai, Y. Li, H. W. Tseng, D. C. Ralph, and R. A. Buhrman, *Science* **336**, 555 (2012).